\documentclass[12pt]{article}
\usepackage{epsfig}
\def\logpm{\ln\frac{1+t}{1-t}}
\def\xb{\bar{x}}
\def\diff{\mbox{d}}
\def\Lbar{\bar{\Lambda}}
\def\lam1{\lambda_1}
\def\GeV{\mbox{ GeV}}

\newcommand{\be}{\begin{eqnarray}}
\newcommand{\ee}{\end{eqnarray}}
\newcommand{\ben}{\begin{equation}}
\newcommand{\een}{\end{equation}}

\newcommand{\ex}[1]{\mbox{e}^{#1}}
\newcommand{\cO}{{\cal O}}

\global\arraycolsep=2pt 
\begin{document}
\makeatletter
\def\fmslash{\@ifnextchar[{\fmsl@sh}{\fmsl@sh[0mu]}}
\def\fmsl@sh[#1]#2{%
  \mathchoice
    {\@fmsl@sh\displaystyle{#1}{#2}}%
    {\@fmsl@sh\textstyle{#1}{#2}}%
    {\@fmsl@sh\scriptstyle{#1}{#2}}%
    {\@fmsl@sh\scriptscriptstyle{#1}{#2}}}
\def\@fmsl@sh#1#2#3{\m@th\ooalign{$\hfil#1\mkern#2/\hfil$\crcr$#1#3$}}
\makeatother
\thispagestyle{empty}
\begin{titlepage}

\begin{flushright}
hep-ph/0101330 \\
CERN-TH/2001-021 \\
TTP01-05 \\
\today
\end{flushright}

\vspace{0.3cm}
\boldmath
\begin{center}
\Large \bf  The determination of
            $V_{ub}$ from inclusive semileptonic $B$ decays
\end{center}
\unboldmath
\vspace{0.8cm}

\begin{center}
{\large Marek Je\.zabek${}^{1,2}$, Thomas Mannel${}^{3,4}$,\\
        Boris Postler${}^3$, Piotr Urban${}^1$} \\
	{\sl ${}^1$Institute of Nuclear Physics, ul. Kawiory 26a, \\
    PL-30055 Cracow, Poland. \\
     ${}^2$Institute of Physics, University of Silesia, ul. Uniwersytecka 4,\\
    PL-40007 Katowice, Poland.\\
     ${}^3$Institut f\"{u}r Theoretische Teilchenphysik, \\
     Universit\"at Karlsruhe,\\ D -- 76128 Karlsruhe, Germany.\\
     ${}^4$CERN-TH, CH-1211 Geneva 23, Switzerland.}
\end{center}

\vspace{\fill}

\begin{abstract}
\noindent
The hadronic mass distribution in semileptonic $B$-meson decays can be
used for extracting the charmless part and thus determining the
$|V_{ub}/V_{cb}|$ ratio. We take into account first-order perturbative
as well as non-perturbative QCD corrections. The sensitivity to model
assumptions is studied and an estimate of the remaining uncertainties is
performed.
\end{abstract}
\end{titlepage}

\section{Introduction}
The CKM matrix element $V_{ub}$ plays an important role in the
determination of the unitarity triangle. The cleanest method to
obtain the absolute value $|V_{ub}|$ is through the measurement of
semileptonic $b \to u$ transitions, which will eventually give
the length of one of the sides of the unitarity triangle.

The possibilities of determining $V_{ub}$ from semileptonic decays
have been studied in detail in the BaBar Workshop \cite{BaBar}.
Here the conclusion was reached that the determination of $V_{ub}$
will be performed by a mixture of different methods.

One of the theoretically
cleanest possibilities is to use inclusive semileptonic
decays. Placing
a cut on the hadronic invariant mass of the final state can in principle
eliminate the charm contribution, which otherwise would be overwhelming.
This cut on the hadronic invariant mass can be implemented
at the asymmetric $B$ factories, making this method experimentally
feasible.

{}From the theoretical side the decay rate, including cuts on the lepton
energy as well as on the hadronic invariant mass of the final state,
can be computed systematically within the framework of the $1/m_b$
expansion, except for certain regions of phase space, where the $1/m_b$
expansion has to be replaced by an expansion in twist. To describe these
regions of phases space one has to introduce a so-called ``shape function'',
which in principle introduces a large hadronic uncertainty.

Another method that has been proposed \cite{Baueretal}
avoids the twist expansion and
relies only on a standard $1/m_Q$ expansion. This method needs a
measurement of the lepton invariant mass spectrum, in which the regions
of phase space where the shape function plays a role are kinematically
suppressed. This method will also allow a clean determination of
$V_{ub}$.

An alternative approach has been put forth in \cite{LLR1,LLR2} where
the factorization into soft, jet and hard sub-processes
\cite{KorchSterm} has been employed to relate the radiative $b$ decays
to the semileptonic ones in a way which explicitly reduces the impact
of the shape function uncertainties. Then a prediction of the ratio
$|V_{ub}/V_{ts}|$  can be obtained
with a good accuracy in a model independent way. This approach is
similar to the one in \cite{MR1}.

Exclusive decays will open a completely different window on $V_{ub}$;
however, in these decays a certain model dependence
seems to be unavoidable, unless lattice data become reasonably precise.

{}From this variety of methods to determine
$|V_{ub}|/|V_{cb}|$, we shall expand in this paper
on the one in which a cut is applied on the
hadronic invariant mass in semileptonic
$B$ decays to filter out the $b \to u$ transitions. The advantage is that
the hadronic invariant mass may be easier to measure, however, this method
involves the shape function and potentially has larger theoretical uncertainties than the inclusive method using the leptonic invariant mass.The method based on the hadronic invariant mass spectrum
has already been discussed in \cite{FLW},
where the main focus was on the perturbative contributions of
order $\alpha_s$ and $\beta_0 \alpha_s^2$. 

The purpose of this paper is to consider this approach in detail and
to try to estimate the uncertainties, including  perturbative
as well as non-perturbative contributions and, in addition, a cut
on the lepton energy. It turns out that the main uncertainties
originate from the heavy
quark mass $m_b$ (or, equivalently, from $\bar\Lambda = M_B - m_b$ where
$M_B$ is the $B$-meson mass) and the strong coupling $\alpha_s$, while
the uncertainties introduced by the shape function are small.

The next section deals with the kinematics. 
Then, in Section \ref{secpert}, we give the radiative corrections for the partonic
process $b \to u \ell \bar{\nu}_\ell$ to order $\alpha_s$. In section
\ref{secshape}
we include the leading-twist non-perturbative effects; this requires the
introduction of the light-cone distribution function for the heavy quark, for
which we use a simple parametrization. We combine the perturbative
and non-perturbative corrections  and
study the uncertainties in the determination of $|V_{ub}|/|V_{cb}|$
in section \ref{Vub}.

\section{Kinematical Relations and Definitions}
We shall first define the kinematic variables for the partonic
process $b\rightarrow u\ell\bar{\nu_\ell}$. Although we also use
results for the semileptonic $b\rightarrow c$ decay rate, the latter
is considered in the partonic framework solely. In fact, we only need
the total rate with a single lower cut on the electron energy for $b \rightarrow c$; we thus 
refer the reader to \cite{jm} for further discussion of the
kinematics of this process.  The initial state $b$ quark has a
momentum $p_b = m_b v$,
where $v$ is the velocity of the $B$ meson. With the momentum transfer
to the leptons $q = k+k'$ the variable $p = q - m_b v$ is the
partonic momentum of the final state. Writing $E_\ell = vk$
for the energy of the lepton, $E_p = vp$ for the partonic energy of the
final state and $p^2$ for the partonic invariant mass, we define the
following re-scaled variables
\begin{equation}
x = \frac{2 E_\ell}{m_b}\, , \qquad x_p = \frac{2 E_p}{m_b} \, , \qquad 
y = \frac{q^2}{m_b^2}\, , \qquad z = \frac{p^2}{m_b^2}\, .
\end{equation}
One of these variables is superfluous when calculating the triple
differential decay rate
and can be substituted by using the relation
\ben
x_p + y -z  = 1 \, .
\een

Using these definitions, we compute the triple differential
partonic rate to order $\alpha_s$ and order $1/m_b$, and split
it into the tree-level term $\Gamma^0$ and the $\cO(\alpha_s)$ correction
$\Gamma^1$:
\begin{equation}
\label{splitting}
\diff\Gamma^{pert}=\diff\Gamma^{parton,0}+\frac{2\alpha_s}{3\pi}
\diff\Gamma^{pert,1} \, .
\end{equation}
The radiative corrections to $\cO(\alpha_s)$ have recently been 
calculated by \cite{hep-ph/9905351}; we have checked their result and
find full agreement with ours. The formulae up to terms $\cO(\alpha_s)$ are
quite tedious and can be found in \cite{hep-ph/9905351}. However, we
present them in the Appendix in a form suitable for our numerical
evaluation.

We shall now discuss the hadronic kinematics. The momentum of the
initial $B$ meson is $M_B v = (m_b + \bar\Lambda) v$, where we have
used the relation between the $B$ meson and the $b$-quark mass
\begin{equation}
M_B = m_b + \bar\Lambda
\end{equation}
which holds to leading order in the $1/m_b$ expansion. Consequently,
the hadronic mass of the
final state is $M_X^2 = (M_B v - q)^2 = (p + \bar\Lambda v)^2
= p^2 + 2 E_p \bar\Lambda + \bar\Lambda^2$ and hence involves both the
partonic invariant mass and the partonic energy. Thus we have
\begin{eqnarray}
\label{hadpart}
\frac{\diff \Gamma^{parton}}{\diff M_X^2 \diff E_l}(m_b)&=&\int\limits_0^1 \diff x
\int\limits_{1-x}^{2-x} \diff x_p \int\limits_{z_{min}}^{z_{max}} \diff z 
\frac{\diff^3 \Gamma^{pert}}{\diff x \diff x_p \diff z}
\times\nonumber\\
&& \delta(M_X^2-m_b^2 z-x_p\Lbar  m_b - \Lbar^2)
   \delta\left( E_l - \frac{m_b x}{2} \right) \,.
\end{eqnarray}
where the kinematic limits of the $z$ integration depend on the cuts
as well as on $x_p$. They are
\be
z_{min} = \left\{
\begin{array}{ll}
0 &\mbox{for } x_p \le 1\, ,\\
x_p-1\,& \mbox{for } x_p > 1\, ,
\end{array}\right.
\ee
and
\ben
z_{max} = (1-x)(x_p+x-1) \, .
\een

%
%
%
%
\section{Perturbative corrections}
\label{secpert}
The analysis we perform requires the knowledge of the triple
differential partonic rate to order $\alpha_s$, as can already be seen
from Eq.~(\ref{hadpart}) as well as from the final formula,
Eq.~(\ref{hadronasconv}). The Born approximation  is proportional to
the delta function of argument $z$,
\ben
\label{partontree}
\frac{\diff^3\Gamma^{pert,0}}{\diff x\,\diff x_p\,\diff
z}=12\Gamma_0\,(2-x-x_p)(x+x_p-1)\delta(z),
\een
where
\ben
\label{gammazero}
\Gamma_0=\frac{G_F^2 m_b^5}{192 \pi^3}.
\een
As is well known, the virtual one-loop correction contains  an infrared
divergence that cancels with the real gluon emission. We shall discuss
in the next section how to include the leading-twist non-perturbative effects
by ``smearing'' over some small window in the hadronic invariant mass.
While choosing an excessively small value for this window
would still yield a meaningless result, a region of size
$\bar\Lambda m_b $ is believed to exist,
where this smearing provides a realistic approximation. Instead of
keeping track of the virtual and real parts, one can now use an
appropriately integrated
distribution, which is subjected to the smearing procedure.
The tree-level
term is rather simple to implement numerically but, on the other hand,
we have found it very useful to perform one integration of the
one-loop correction analytically. On inspection of
Eq.~(\ref{hadronasconv}) in conjunction with Eq.~(\ref{hadpart}), it
is clear that the following integral is of use:

\begin{equation}
F(x,x_p,z)=\int_{z_{min}}^z \diff z' \frac{\diff \Gamma
  ^{pert,1}}{\Gamma_0\, \diff x \diff x_p \diff z'}.
\end{equation}

This function is a lengthy and tedious expression, and it is
given in the Appendix.

\section{Leading-twist non-perturbative corrections}
\label{secshape}
It has been shown that the leading-twist non-perturbative corrections
can be implemented at tree level
by redefining the heavy-quark mass and a
subsequent convolution with a so-called shape function
\cite{hep-ph/9311325,
hep-ph/9312311,hep-ph/9402288,hep-ph/9312359,hep-ph/9402225}.
This convolution corresponds to an integration over the light-cone
variable $k_+$, namely
\ben \label{conv}
\diff\Gamma^{hadron} = \int\limits^{\bar\Lambda}_{-m_b}
\diff\Gamma^{parton}(m_b + k_+)
\,f(k_+)\,\diff k_+
\een
in $\diff\Gamma^{HQET}$.
Although this formula is quite suggestive \cite{mr3}, it has been
shown recently that it contains,in fact, spurious contributions of
sub-leading twist \cite{BLM}. Furthermore, once radiative corrections
are implemented, this simple convolution formula will probably no 
longer hold \cite{LukeLigeti}. However, there is no fully
consistent way  yet to include radiative corrections into this convolution,
and hence we proceed in a naive way as suggested in \cite{mr3}.

This naive way is to actually use the convolution formula (\ref{conv})
also beyond tree level, which is at least as consistent as using the
ACCMM model \cite{ACCMM} beyond tree level,
which is common practice. In fact,
the connection between the ACCMM model and the shape function formalism
has been pointed out in \cite{BIgiACCMM}.

It is convenient to change the variable of integration according to
\ben
\label{hadronasconv}
\diff\Gamma^{hadron} = \int\limits^{M_B}_{0}\diff\Gamma^{parton}(m^*)
\,f(m^*)\,\diff m^*
\een
with $m^* = m_b +k_+$.

The shape function is a nonperturbative function which  has to
be determined either from experiment or by some model. A few
relations are known for the moments of the shape function
$\langle k_+^n\rangle$:
\ben
\label{moments}
\langle k_+^0\rangle = 1 \, ,\quad
\langle k_+^1\rangle = 0 \quad \mbox{and}\quad
\langle k_+^2\rangle = -\frac13 \lambda_1 \quad 
\een

For our study we shall use an ansatz for the
shape function. Taking a simple three parameter function
\cite{hep-ph/9805303}
\ben
\label{shapefunction}
f(x) = N\, \ex{c\, x}\, (1-x)^a  \, ,  \quad
x=1 - \frac{M_B - m^*}{\bar\Lambda} \, , 
\een
we can express the parameters of this ansatz in terms of the
HQET parameters (\ref{moments})
\ben
N= \frac{c^c}{\bar\Lambda\ex{c}\Gamma(c)} , \quad
a= c-1   , \quad
c= -\frac{3\bar \Lambda^2}{\lambda_1}\, .
\een
We shall discuss the dependence of our results on this ansatz
in the next section.

\section{The measurement of $|V_{ub}|$}
\label{Vub}
For the measurement of $|V_{ub}|$ we propose
to study semileptonic $B$ decays with certain cuts.
The first cut is on the lepton energy $E_l$, which is mainly given by
the experimental limitations on detecting electrons with small momenta.
The second cut is on the hadronic invariant mass $M_X$ of the final
state, which serves to suppress charm. We define the semileptonic rates
including cuts as
\begin{equation}
\Gamma (M_{cut}^2,E_{cut})=\int\limits_0^{M_{cut}^2}\diff M_X^2
\int\limits_{E_{cut}}^{\frac{M^2_B-M_X^2}{2\,M_B}} \diff E_l \, \frac{\diff 
\Gamma(B\rightarrow
  X_u e\bar{\nu_e})}{\diff M_X^2 \diff E_l},
\end{equation}
Clearly the region $0 < M_X < M_D$ is dominated by $b \to u$ transitions,
and we thus use in this region the expressions for $b \to u$ decays only.

We shall normalize everything to the rate with no cut on the hadronic
invariant mass, and thus obtain, for the ratio $r$:
\begin{equation}\label{rratio}
\frac{|V_{ub}|^2}{|V_{cb}|^2}r=
\frac{\Gamma(M_{cut}^2, E_{cut})}{\Gamma(M_B,E_{cut})},
\end{equation}
where in the denominator we can safely take into account the $b \to c $
channel only, the $b \to u$ contribution being only
about 1\%.

For the charmless decay rate entering the  numerator, we
include the results discussed above. The perturbative corrections
are taken into account to one loop
\cite{jm}, while the non-perturbative
ones are included to leading twist.
The $b \to c$ decay rate in the denominator is evaluated including
${\cal O}(\alpha_s)$ corrections.

Unlike the approach advertised in \cite{Baueretal} and \cite{Neubertnew},
the shape of the hadronic invariant mass spectrum depends on the shape
function, for which we use the parametrization (\ref{shapefunction}). However,
as it will turn out, the precise form of the shape function is irrelevant
for the ratio (\ref{rratio}), the main sources of uncertainties are
the value of the strong coupling and that of the quark masses. The 
latter may be replaced by $\Lbar$, we thus use
$m_b = \overline{M}_B - \Lbar$ and $m_c = \overline{M}_D - \Lbar$, where
$\overline{M}$ denotes the spin-averaged meson mass.

The dependence on the shape function can be studied by checking the
sensitivity of our result with respect to higher moments of this
function. Since we have not included any $1/m_b^2$ terms in our
calculation the ratio (\ref{rratio}) should be reasonably insensitive
already to the second moment of the shape function, which is given in
terms of the kinetic energy operator $\lambda_1$ in (\ref{moments}).
This forces us to make $M_{cut}$ as large as possible.  

We have examined the ratio defined in Eq.~(\ref{rratio}), allowing the
variations $0.4 \GeV \le \Lbar \le 0.75 \GeV$, $0.2 \le \alpha_s \le 0.3$
and $-0.6 \GeV^2 \le \lam1 \le -0.1 \GeV^2$.

\begin{figure}
\begin{center}
\epsfig{file=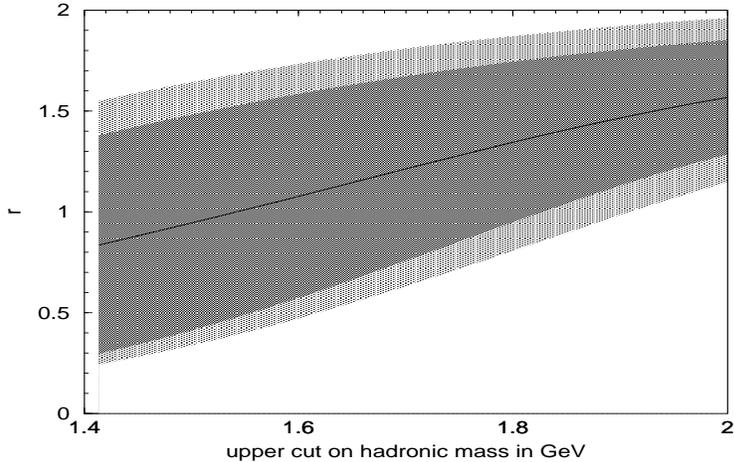,width=300pt,height=200pt}
\caption{Ratio $r$ as a function of the hadronic mass cut, with (dark)
or without (light) correlation between the strong coupling constant
and the pole mass}
\label{rband}
\end{center}
\end{figure}

The main uncertainty is induced by $\alpha_s$ and $\Lbar$, where $\Lbar$ 
is equivalent to the heavy quark mass. It has been argued that these two
quantities are correlated; the size of the radiative corrections depends
on the particular choice of the mass.

Using the pole mass scheme, it has been shown that the radiative corrections
are large and the perturbation series converges very slowly. Treating
both the pole mass and the perturbative contributions
independently, the uncertainties of the quantities would simply add, 
leaving us with a large (and certainly overestimated)
uncertainty. In this case we get 
(see the lighter band in Fig. \ref{rband}):
\ben
1.15 \le r \le 1.96 \quad \mbox{at } M^2_{cut}=4 \GeV ^2
\een
where we have used $m_b^{pole} = (4.75 \pm 0.15) $ GeV and $\alpha_s$ 
between 0.2 and 0.3.
In this way we obtain a theoretical  uncertainty
of $25\%$ in the determination of the
ratio.

Another option is to switch to a short-distance mass definition, such as
$m_b^{\overline{MS}}$ by replacing
\ben
m_b^{pole}=m_b^{\overline{MS}}\left( 1 +\frac{4}{3\pi}\alpha_s\right),
\een
which reduces the size of the coefficients of the perturbation series,
and the pertrubative uncertainties thus become smaller. Using a recent
value for $M_b^{\overline{MS}}$ \cite{Neubertnew}
\ben
m_b^{\overline{MS}}=(4.25\pm 0.08) \mbox{GeV},
\een
we arrive at an estimate for $r$ with a smaller uncertainty (the darker band in Fig. \ref{rband})
\ben
1.28 \le r \le 1.85 \quad \mbox{at } M^2_{cut}=4\GeV^2.
\een

In order to display the dependence on the input parameters, we
choose the ``average'' values of the three parameters:
\begin{equation}
\Lbar^{aver}=0.55 \GeV,\qquad \alpha_s^{aver}=0.25,\qquad
\lam1^{aver}=-0.35 \GeV^2,
\end{equation}
and obtain up to linear terms in the variations
\ben
r=1.58-0.86\frac{\Delta \Lbar}{\Lbar^{aver}}-0.26\frac{\Delta
 \alpha_s}{\alpha_s^{aver}}+0.06\frac{\Delta\lam1}{\lambda_1^{aver}} .
\een
This explicitly shows that the dependence on $\lambda_1$, which is
the second moment of the shape function, is weak. The dependence on
even higher moments is expected to be  further suppressed by inverse
powers of the heavy quark mass.

In conclusion, our best estimate for $r$ is
\ben
r=1.57\pm 0.3\quad ,
\een
which corresponds to a
theoretical uncertainty in the determination of
 $ | V_{ub} / V_{cb} | $ of about ten percent
\begin{equation}
\left(\frac{\Delta| V_{ub} / V_{cb} |}{| V_{ub} / V_{cb} |} \right)_{theor}
\approx 10 \%
\end{equation}

\section{Conclusions}
\label{secconclusions}
We have performed a detailed analysis of one of the possibilities
to obtain $V_{ub}$ from inclusive semileptonic $B$ decays by placing
a cut on the hadronic invariant mass to get rid of the charm background.
This method has been criticized since it depends on the shape
function, which describes the endpoint of the hadronic invariant mass
spectrum. This function is not very well known and hence it has to be
modelled, which will introduce some systematic uncertainty.
However, integrating over the window in hadronic invariant masses
relevant to $b \to u$ transitions, we have shown that the dependence
on the shape function is much smaller than the uncertainties induced
by the quark mass and by the truncation of the perturbative series.

The the ratio between the semileptonic rates including a cut on the
hadronic invariant mass and the semileptonic rate without a cut yields
$|V_{ub}/V_{cb}|$ up to a quantity $r$, which we have computed
in leading twist approximation and to order $\alpha_s$. Based on
our calculations the uncertainty in this quantity is $20 \%$ leaving us
with a $10\%$ theoretical uncertainty in the determination of $V_{ub}$.
This method is thus one of the cleanest possible
to obtain $V_{ub}$ at the ongoing $B$-factory experiments.

\section*{Acknowledgements}
TM and BP acknowledge support from the DFG Graduiertenkolleg
``Elementarteilchenphysik an Beschleunigern''and from the
DFG Forschergruppe ``Quantenfeldtheorie, Computeralgebra und
Monte Carlo Simulationen''. TM also acknowledges support from
the BMBF. This work is partly supported by the KBN grants 2P03B05418 and 5P03B09320 and by the European Commission 5th Framework contract HPRN-CT-2000-00149.
PU would like to thank the Polish--French Collaboration within IN2P3 through Annecy.

\section*{Appendix}
\renewcommand{\theequation}{A.\arabic{equation}}
\setcounter{equation}{0}
We present here the convoluted spectra in some more detail, 
showing in particular a convenient way of calculating the contribution
from the terms proportional to $\alpha_s$.
The convolution of the partonic spectra results in the leading-twist
approximated rate of $B\rightarrow X_u l \bar{\nu}$ decay.
 The resulting distribution can be written in the following form:
\begin{eqnarray}
\frac{\diff \Gamma}{\diff M_X^2 \diff E_l}&=&\int_0^1 \diff x
\int_{1-x}^{2-x} \diff x_p \int_{z_{min}}^{z_{max}} \diff z
\int_{-\infty}^{M_B} \diff m^*
f(m^*) \frac{\diff \Gamma^{parton}}{\diff x \diff x_p \diff
z}\times\nonumber\\
&& \delta(M_X^2-m^{*2}z-x_p\Lbar^* m^* - \Lbar^{*2}) \delta(E_l -
 \frac{m^* x}{2}).
\end{eqnarray}
The partonic rate can be split into the tree level term and the ${\cal
  O}(\alpha_s)$ correction,
\begin{equation}
\diff\Gamma^{parton}=\diff\Gamma^{parton,0}+\frac{2\alpha_s}{3\pi}
\diff\Gamma^{parton,1},
\end{equation}
and the convoluted rate can then be divided up accordingly. Using the
delta functions, the integrations involved in these rates can be
simplified, yielding
\begin{eqnarray}
&&\int_{E_{cut}}^{M_B/2} \diff E_l \int_0^{M_{cut}^2}\diff M_X^2
 \frac{\diff \Gamma^{(0)}(B\rightarrow
  X_u e\bar{\nu_e})}{\diff M_X^2 \diff E_l}=\nonumber\\
&&\int_{E_{cut}}^{M_B/2}\diff E_l \int_{2E_l/M_B}^1 \diff x
 \int_{y_{min}}^x\diff y\,
\frac{2}{x}f\left( \frac{2E_l}{x} \right)\frac{\diff\Gamma^{parton,0}}{
 \diff x \diff y},
\end{eqnarray}
\begin{equation}
y_{min}=\max \left\{ 0,
\frac{1-M_{cut}^2-(M_B-2E_l/x)^2}{(2E_l/x)(M_B-2E_l/x)}\right\}
\end{equation}
for the tree-level contribution, while the $\alpha_s$ correction
contributes
\begin{eqnarray}\label{CorrConv}
&&\int_{E_{cut}}^{M_B/2} \diff E_l \int_0^{M_{cut}^2}\diff M_X^2
 \frac{\diff \Gamma^{(1)}(B\rightarrow
  X_u e\bar{\nu_e})}{\diff M_X^2 \diff E_l}=\nonumber\\
&&\int_{E_{cut}}^{M_B/2}\diff E_l\int_{2E_l/M_B}^1\diff x
\int_{1-x}^{2-x} \diff x_p
 \int_{z_{min}}^{z_{max}} \diff z\, \frac{2}{x} f\left(\frac{2E_l}{x}\right)
 \frac{\diff\Gamma^{parton,1}}{\diff x \diff x_p \diff z},
\end{eqnarray}
where
\begin{equation}
z_{min}=\left\{
\begin{array}{cc}
  0,&{\mbox for\ } x_p \le 1,\\
  x_p-1&{\mbox for\ } x_p > 1,
\end{array}\right.
\end{equation}
and
\begin{equation}
z_{max}=\min\left\{ \frac{M_{cut}^2-x_p (M_B-2E_l/x)-(M_B-2E_l/x)^2}{2E_l/x},
(1-x)(x_p+x-1)\right\}.
\end{equation}
With these formulae, it is easy to numerically convolute the Born
approximated rate. However, it is useful to eliminate one integral from
the convolution of the $\alpha_s$ term. To this end,
the integral over $z$ in Eq. (\ref{CorrConv}) has been performed
analytically, so that we use the prime function $F(x,x_p,z)$ defined as
\begin{equation}
F(x,x_p,z)=\int_{z_{min}}^z \diff z' \frac{\diff \Gamma
  ^{parton,1}}{\Gamma_0\, \diff x \diff x_p \diff z'},
\end{equation}
where $\Gamma_0$ is defined in $(\ref{gammazero})$. Then,
 \begin{equation}
F=\left\{
\begin{array}{ll}
        F_1(z)-F_1(0)+F_2(z)+F_3,&x_p<1,\nonumber\\
        F_1(z)-F_1(z_{min})+F_2(z)-F_2(z_{min})&x_p>1,
\end{array}
\right.
\end{equation}
where
\begin{equation}
z_{min}=x_p-1,
\end{equation}
and
\begin{eqnarray}
F_1&=&\frac{1}{8}\left[{1\over t}\logpm [x_p^2-4(x-1+x_p/2)^2/t^2]
+8(x-1+x_p/2)^2/t^2\right]\nonumber \\
&&\times(36-39 x_p + 12 x_p^2 - 3 x_p^3/4)+{1\over
t}\logpm(x-1+x_p/2)[48 - 78 x_p\nonumber \\
&& + 45 x_p^2 - 21 x_p^3/2 + 3x_p^4/4
 + (x-1+x_p/2)  (  - 42 - 15 x_p + 27x_p^2 \nonumber\\
&&- 21/8 x_p^3 )]
        +(A_1 t + A_2)\logpm
        +A_3\log(1-t)
        +A_4,\\
F_2&=& 12[2\ln^2 z-(8\ln x_p-7)\ln z]v(\xb-x_p),\\
F_3 &=&(x_p-\xb)\left\{ 24\ln x_p \left(-1+5v-4v\ln
x_p\right)\right.\nonumber \\
&&\left.
        +v[-16\pi^2-60-48 {\mbox{Li}_2}(1-x_p)]\right\}.
\end{eqnarray}
In the above formulae,
\begin{eqnarray}
A_1&=&(x_p-\xb)  ( x_p z \xb/2 + 21 x_p \xb/2 - 6 x_p
         v - 12 x_p^2 \xb + 3 x_p^2 v + 7x_p^3\xb/4 )\nonumber\\
&&+ ( 2 x_p z v - 9 x_p z/2 - 9/10x_pz^2 -
         x_p^2 z v + 7x_p^2z + 12x_p^2 v+ 15x_p^2/2 \nonumber \\
&&- 21x_p^3z/20 + 17x_p^3 v/2 - 105/4x_p^3 - 5/4x_p^4 v + 53/4
         x_p^4\nonumber \\
&& - 219/160x_p^5 ),\\
A_2&=&(x_p-\xb)  (  -36 x_p (1-x_p x) + 6x_p^2v + 4x_p^3\xb +
12\xb - 96v\ln2 + 48v )\nonumber \\
&&- 24x_p + 48x_p^2 + 16x_p^3v - 60
         x_p^3 - 2x_p^4v + 26x_p^4 - 12/5x_p^5,\\
A_3&=&2A_2+96(x_p-\xb)(1+\xb-x_p)\ln(1+t),\\
A_4&=&  (x_p-\xb)  ( 5x_p z\xb - 24z\xb + 12zv )
       + 32x_pzv- 81x_pz - 21/5x_pz^2 \nonumber \\
&& - 4x_p^2zv+ 40x_p^2z - 81/20x_p^3z + 36zv + 12z - 6z^2v + 24
         z^2,
\end{eqnarray}
and
\begin{equation}
\xb=1-x,\qquad v=2-x-x_p,\qquad t=\sqrt{1-4z/x_p^2}.
\end{equation}


\begin{thebibliography}{10}
\bibitem{BaBar}
P. F. Harrison and H. R. Quinn, eds.,The BaBar physics book:
Physics at an asymmetric B  factory, Papers from Workshop on Physics
at an Asymmetric B Factory (BaBar Collaboration Meeting), Rome, Italy,
11-14 Nov 1996, Princeton, NJ, 17-20 Mar 1997, Orsay, France, 16-19 Jun
1997 and Pasadena, CA, 22-24 Sep 1997.

\bibitem{Baueretal}
C.~Bauer, Z.~Ligeti and M.~Luke, Phys.Lett. {\bf B479} (2000) 395-401; hep-ph/0002161.

\bibitem{LLR1}
A.~K.~Leibovich, I.~Low and I.~Z.~Rothstein, Phys. Rev. {\bf D62}
(2000) 014010.

\bibitem{LLR2}
A.~K.~Leibovich, I.~Low and I.~Z.~Rothstein, Phys. Lett. {\bf B486}
(2000) 86.

\bibitem{KorchSterm}
G.~P.~Korchemsky and G.~Sterman, Phys. Lett. {\bf B340} (1994) 96.

\bibitem{MR1}{T. Mannel, S. Recksiegel,
              Phys.\ Rev.\ {\bf D60} (1999) 114040.}  


\bibitem{FLW}
A.~Falk, Z.~Ligeti and M.~B.~Wise, Phys. Lett.{\bf B406} (1997) 225.

\bibitem{jm}
M. ~Je\.zabek and L. ~Motyka, Acta Phys. Polon. {\bf B 27}, 3603
(1996); Nucl. Phys. {\bf B 501}, 207 (1997).

\bibitem{hep-ph/9905351}
F.~DeFazio and M.~Neubert,
\newblock JHEP {\bf 06}, 017 (1999), hep-ph/9905351.

\bibitem{hep-ph/9311325}
M.~Neubert,
\newblock Phys. Rev. {\bf D49}, 3392 (1994), hep-ph/9311325.

\bibitem{hep-ph/9312311}
M.~Neubert,
\newblock Phys. Rev. {\bf D49}, 4623 (1994), hep-ph/9312311.

\bibitem{hep-ph/9402288}
T.~Mannel and M.~Neubert,
\newblock Phys. Rev. {\bf D50}, 2037 (1994), hep-ph/9402288.

\bibitem{hep-ph/9312359}
I.~I. Bigi, M.~A. Shifman, N.~G. Uraltsev, and A.~I. Vainshtein,
\newblock Int. J. Mod. Phys. {\bf A9}, 2467 (1994), hep-ph/9312359.

\bibitem{hep-ph/9402225}
I.~Bigi, M.~Shifman, N.~Uraltsev, and A.~Vainshtein,
\newblock Phys. Lett. {\bf B328}, 431 (1994), hep-ph/9402225.

\bibitem{mr3}
T.~Mannel, S.~Recksiegel, hep-ph/0009268, in print in Phys. Rev. {\bf D}.

\bibitem{BLM}{C. Bauer, M. Luke, T. Mannel, 
CERN preprint CERN-TH-2001-027.} 

\bibitem{LukeLigeti}
Z. Ligeti, M. Luke, in preparation.

\bibitem{ACCMM}
G. Altarelli et al., Nucl. Phys. {\bf B208} (1982) 365.

\bibitem{BIgiACCMM}
I. Bigi et al., Phys. Lett. {\bf B328} (1994) 431.

\bibitem{hep-ph/9805303}
A.~L. Kagan and M.~Neubert,
\newblock Eur. Phys. J. {\bf C7}, 5 (1999), hep-ph/9805303.

\bibitem{Neubertnew}
M.~Neubert, JHEP {\bf 0007} (2000) 022.
\end{thebibliography}
\end{document}